# Chiral Symmetry Breaking in Abelian-Projected SU(2) Lattice Gauge Theory

R.M. Woloshyn

*TRIUMF, 4004 Wesbrook Mall, Vancouver, B.C. V6T 2A3*

## Abstract

Chiral symmetry breaking parameters are calculated in quenched SU(2) lattice gauge theory and with Abelian gauge fields projected in maximal Abelian gauge and in field strength gauge. Maximal Abelian gauge projected fields lead to chiral condensate values which are quite similar to those of the full nonabelian theory. Pseudoscalar and vector meson correlators are calculated and found to be reproduced by the use of maximal Abelian gauge fields for small quark masses. In field-strength gauge, Abelian-projected fields give a chiral condensate which closely resembles the results of strongly coupled ($\beta < 1$) gauge theory: the chiral condensate is insensitive to $\beta$ and quark mass and hence violates scaling badly.

# I. INTRODUCTION

The role of Abelian monopoles [1,2] in describing confinement in nonabelian theories has been the subject of numerous investigations using lattice field theory [3–7]. Up to now these studies have been restricted to the gauge field sector. Such questions as the value of the string tension in Abelian-projected fields [3], an Abelian-monopole order parameter for the finite temperature deconfinement transition [4,5] and the possible form for effective Abelian gauge field actions have all been addressed [5,7]. In this paper we take the study of the Abelian monopole mechanism in a new direction. The response of fermions to Abelian-projected fields is calculated within the framework of a quenched SU(2) gauge theory focussing on the comparison of chiral symmetry breaking in the projected and the original SU(2) theories.

The lattice formulation of Abelian projection was developed in Ref. [8,9]. Several gauge fixing conditions have been studied and it has been found that the so-called maximal Abelian gauge [9] gives the clearest view of the Abelian monopole mechanism. In this gauge, the quantity (for SU(2) gauge field links $U_\mu(x)$)

$$\mathcal{R} = \sum_{x,\mu} \mathrm{Tr}(\sigma_3 U_\mu(x) \sigma_3 U_\mu^\dagger(x)) \,, \tag{1}$$

is maximized. This means that, on average, the links are as diagonal as possible. The Abelian-projected field produced in this gauge is relatively smoother than that found with gauge conditions such as field-strength diagonalization which can be imposed locally, that is, point by point on the lattice. This is manifested by the fact that the density of Abelian monopoles (related to singularities in the gauge transformation imposing the gauge condition) is typically an order of magnitude smaller in maximal Abelian gauge than in "local" gauges [10]. More detailed investigation shows that the surplus of Abelian monopoles in "local" gauges is largely made up of short-distance correlated monopole-antimonopole pairs[11].

Another way to summarize the difference between Abelian projection in maximal Abelian gauge and in a local gauge fixing, such as field-strength diagonalization, is to say that the Abelian links produced in field-strength gauge are much more random than those in the more restrictive maximal Abelian gauge. In other words, the effective U(1) theory describing a field-strength gauge projection is at a much stronger coupling than the U(1) theory describing a maximal Abelian projection. This is clear from the calculation of the string tension [3]. It has been found that Wilson loops constructed from Abelian-projected links in maximal Abelian gauge closely match the original SU(2) Wilson loops and so the string tension of the projected theory is very similar to the string tension of the full theory. This has been dubbed Abelian dominance [3]. On the other hand, the Wilson loops from, for example, field-strength gauge projection are relatively suppressed and the corresponding string tension is much larger than in the unprojected SU(2) theory.

The behaviour of quantities in the continuum limit is an important concern in lattice field theory. Previous studies have shown that the density of monopoles for maximal Abelian gauge in physical units is consistent with being constant as a function of $\beta$ (scaling behaviour)[11,12]. In contrast, the density of elementary ($1^3$) monopoles in local gauges is roughly constant in lattice units, i.e., scaling is badly violated.

The Abelian monopole mechanism in maximal Abelian gauge provides a very attractive picture of confinement of static colour charges as due to a plasma of monopoles and antimonopoles in the QCD vacuum. It is natural to ask what happens if one considers light quarks interacting via the Abelian part of the gauge field links. To this end we have carried out a numerical study of quenched SU(2) lattice gauge theory with staggered fermions. The main part of the work deals with chiral symmetry breaking which is of central importance in understanding the low-energy properties of QCD. Of course chiral symmetry breaking in SU(2) lattice gauge theory has been calculated before. What is new here is that it was carried out using Abelian-projected fields.

The expectation value of staggered fermion fields $< \overline{\chi}(0)\chi(0) >$ was calculated for nonzero fermion bare mass values. As a function of mass this expectation value evaluated using maximal Abelian gauge projected links differs from that calculated in the original SU(2) theory. However, the chiral condensate ($< \overline{\chi}(0)\chi(0) >$ extrapolated to zero fermion mass) is quite similar in the two calculations. Quite remarkably it was also found that the chiral condensate calculated with maximal Abelian projected links obeys perturbative scaling almost exactly.

In addition to chiral symmetry breaking, meson correlators for the spin 0 and 1 channels were also calculated. The qualitative trend is that the maximal Abelian gauge links can reproduce the shape of the pseudoscalar correlator (i.e., the pseudoscalar meson mass) quite well for all values of the (bare) quark mass. On the other hand, the splitting between pseudoscalar and vector channels is observed to decrease faster with increasing quark mass in the maximal Abelian calculation than in the SU(2) case. This together with the different dynamical mass generation evident at nonzero fermion mass suggests the qualitative conclusion that maximal Abelian projected fields reproduce the long distance behaviour of SU(2) theory very well but some short distance effects are washed out in the Abelian projection.

From string tension calculations it is anticipated that Abelian links in field-strength gauge would produce results reminiscent of strongly coupled U(1) gauge theory. Indeed this is what is found for the expectation value $< \overline{\chi}(0)\chi(0) >$. With field-strength gauge projected links it is larger than the SU(2) value and essentially independent of mass or coupling constant.

## II.   METHOD

The Wilson plaquette action

$$S = \beta \sum_{x,\mu<\nu} (1 - \frac{1}{2}\mathrm{Tr}U_{\mu\nu}(x)) \tag{2}$$

is used with periodic boundary conditions in all directions. A heatbath Monte Carlo algorithm was used to construct gauge field configurations.

For an SU(2) gauge theory the Abelian projection is particularly simple. Each link is factorized

$$U_\mu(x) = w_\mu(x)u_\mu(x) \tag{3}$$

where $w$ and $u$ have the form

$$w = w^0 I + i(\sigma_1 w^1 + \sigma_2 w^2)\,, \tag{4}$$

and

$$u = u^0 I + i\sigma_3 u^3\,, \tag{5}$$

where the $\sigma$ s are the Pauli matrices. The diagonal factor $u$ will be referred to as the Abelian-projected field.

In this paper two possible gauge fixing conditions are considered before the projection of Eq. (3) is carried out. One is the so-called maximal Abelian gauge [5] defined in a lattice theory, as the gauge in which the quantity $\mathcal{R}$ in Eq. (1) is maximized. This condition is nonlocal. Under a gauge transformation $G(x)$ each term in $\mathcal{R}$ involves $G$s from neighbouring sites. In practice $\mathcal{R}$ is maximized iteratively, repeatedly sweeping through the lattice maximizing $\mathcal{R}$ locally by solving for $G(x)$ at some particular $x$ keeping the neighbouring sites fixed.

For purposes of comparison, calculations were also done in a local gauge which can be defined by the condition that some adjoint operator $\Phi(x)$ be diagonal [2,4], that is, the gauge transformation $G(x)$ is determined (site by site) by

$$\Phi(x) \rightarrow G(x)\Phi(x)G^\dagger(x) = \text{diagonal}. \tag{6}$$

Here we choose $\Phi(x)$ to be the (lattice) field strength component $U_{12}(x)$.

The focus of this paper is on how fermions behave in Abelian-projected fields, that is, what happens if in a sample of SU(2) gauge configurations the links $U_\mu(x)$ are replaced by the diagonal parts $u_\mu(x)$ calculated in some gauge. Of particular interest is spontaneous chiral symmetry breaking which is a basic feature of QCD. The chiral symmetry properties are most easily studied using staggered fermions. The action is

$$S_f = \frac{1}{2}\sum_{x,\mu}\eta_\mu(x)\Big[\overline{\chi}(x)U_\mu(x)\chi(x+\hat{\mu}) - \chi(x+\hat{\mu})U_\mu^\dagger(x)\chi(x)\Big] + \sum_x m\overline{\chi}(x)\chi(x)\,, \tag{7a}$$

$$\equiv \overline{\chi}\mathcal{M}(\{U\})\chi\,, \tag{7b}$$

where $\overline{\chi}, \chi$ are the single-component staggered-fermion fields and $\eta_\mu(x)$ is the staggered-fermion phase [13]. Antiperiodic boundary conditions were used for the fermion fields in all directions.

The chiral symmetry order parameter $<\overline{\chi}\chi>$ is determined from the inverse of the fermion matrix $\mathcal{M}$ by

$$<\overline{\chi}\chi> = \frac{1}{V} < \text{Tr}\mathcal{M}^{-1}(\{U\}) > \tag{8}$$

where $V$ is the lattice volume and the angle brackets denote the gauge field configuration average. For each configuration $\text{Tr}\mathcal{M}^{-1}(\{U\})$ was calculated using a random source method [14,15] with 12 Gaussian random sources.

In addition to chiral symmetry breaking, meson correlators for the spin 0 and spin 1 channels were also calculated. These correlators can be constructed from local bilinears of the staggered fields and after integration of the fermion fields they take the form

$$g_0(t) = \sum_{\vec{x}} \text{Tr}\{\mathcal{M}^{-1}(\vec{x}, t; 0)[\mathcal{M}^{-1}(\vec{x}, t; 0)]^\dagger\} \qquad (9a)$$

and

$$g_1(t) = \sum_{\vec{x}} [(-1)^{x_1} + (-1)^{x_2} + (-1)^{x_3}]\text{Tr}\{\mathcal{M}^{-1}(\vec{x}, t; 0)[\mathcal{M}^{-1}(\vec{x}, t; 0)]^\dagger\} \qquad (9b)$$

for spin 0 and 1 respectively.

## III.   RESULTS

The chiral symmetry breaking study was carried out on a $14^4$ lattice for a range of $\beta$ values from 2.3 to 2.5. At each $\beta$ the gauge field was equilibrated for 4000 heatbath Monte Carlo sweeps. Then $< \overline{\chi}\chi >$ was calculated for 20 configurations separated from each other by 300 sweeps. The calculation was done for values of the staggered fermion mass $ma$=0.05 to $ma$=0.3 ($a$=lattice spacing).

For each gauge field the $SU(2)$ links were replaced by their Abelian projection determined after imposing the maximal Abelian gauge or the field-strength gauge and $< \overline{\chi}\chi >$ was recalculated. The results in lattice units are tabulated in Table I. A sample of results ($\beta = 2.3$ and $\beta = 2.5$) are shown in Figs. 1 and 2.

The results using maximal Abelian gauge links are similar to those obtained in the full $SU(2)$ theory, and the similarity increases at smaller masses. On the other hand, with field-strength gauge, $< \overline{\chi}\chi >$ is much larger and almost independent of mass. This is the behaviour typically observed in the strong coupling region ($\beta < 1$) and reinforces the notion that with field strength diagonalization the resulting Abelian links are very random.

To determine if chiral symmetry is spontaneously broken one needs to extrapolate $< \overline{\chi}\chi >$ to zero mass after the infinite volume limit is taken. In practise, finite volume calculations in some mass window, where it is believed finite volume effects are small, are used along with a hypothesis for the extrapolation function to estimate $< \overline{\chi} \chi >$ at zero mass. To determine a suitable mass window some calculations on $12^4$ and $16^4$ lattice were also done. Even at our smallest mass, $ma = 0.05$, essentially no dependence in $< \overline{\chi} \chi >$ on lattice volume was observed.

The choice of extrapolation procedure is more problematic. For simplicity the method of Billoire et al.[16] is adopted. The quantity $< \overline{\chi} \chi >$ is expanded in powers of mass, keeping the first three terms

$$< \overline{\chi} \chi > (m) = < \overline{\chi} \chi >_0 + < \overline{\chi} \chi >_1 m + < \overline{\chi} \chi >_2 m^2 . \tag{10}$$

The coefficients are determined by fitting to computed values. To get some handle on the systematic uncertainty of this procedure, fits to different sets of the mass points were done. The resulting zero mass values $< \overline{\chi} \chi >_0$ are given (in lattice units) in Table II for two fits; one to all mass points (6-point fit) and another to the mass points $ma$=0.1 to 0.3 (5-point fit). The fitted extrapolation functions are also plotted in Figs. 1 and 2 for $\beta$=2.3 and 2.5 respectively.

What is seen is that Eq. (10) does not give a completely accurate representation of $< \overline{\chi} \chi > (m)$ for the SU(2) theory. The 5-point fit yields values of $< \overline{\chi} \chi >_0$ which are systematically higher than those resulting from including the $ma$=0.05 point. The function $< \overline{\chi} \chi > (m)$ calculated using maximal Abelian links is fit very well by Eq. (10).

Next we consider the scaling of the chiral condensate $< \overline{\chi} \chi >_0$ as a function of $\beta$. Perturbatively it is expected that [16,17]

$$(< \overline{\chi} \chi >_0)^{\frac{1}{3}} \propto \beta^{507/968} e^{-\frac{3}{11}\pi^2 \beta} , \tag{11}$$

where this behaviour incorporates the two-loop result for the lattice spacing a$(\beta)$ and the anomalous dimension of $< \overline{\chi} \chi >$. To check for perturbative scaling the ratio of ratios $r_{\text{P}}(\beta)/r_{\text{P}}(\beta = 2.3)$ with

$$r_{\text{P}}(\beta) = (< \overline{\chi}\chi >_0)^{\frac{1}{3}}/(\beta^{507/968}e^{-\frac{3}{11}\pi^2\beta}) \tag{12}$$

is plotted in Fig. 3. The maximal Abelian gauge results are remarkably consistent with perturbative scaling. Due to the systematic uncertainty of the fit, no definite conclusion can be drawn about the full SU(2) calculation. The 6-point fit results are certainly not consistent with perturbative scaling. The trend is that $< \overline{\chi}\chi >_0$ decreases more rapidly than expected from Eq. (11).

It is well known that other quantities do not obey perturbative scaling for our values of $\beta$. We have also checked scaling compared to the string tension. Figure (4) shows the ratio $r_{st}(\beta)/r_{st}(\beta = 2.3)$ for $r_{st}(\beta) = (< \overline{\chi}\chi >_0)^{\frac{1}{3}}/\sqrt{K}$ where $K$ is the string tension. For the string tension the results of Michael and Teper [18] were used. Since the maximal Abelian gauge results follow perturbative scaling they do not scale with $\sqrt{K}$. The SU(2) results are again inconclusive. The 6-point fit results scale better with the string tension than with Eq. (11) but scaling of the 5-point fit results is worse. In any case the scaling violation are moderate, 10% or less, which is comparable to what other studies have found [16,17,19].

Since the chiral order parameter calculated using the maximal Abelian gauge projected links is approximately the same as that in the full SU(2) theory, it is natural to ask about hadron properties. To get some qualitative information about this, meson correlators for the pseudoscalar and vector channels, Eq. (9) were also calculated. This was done on a $12^3 \times 20$ lattice using 30 configurations separated by 300 heatbath Monte Carlo sweeps. The correlators constructed using maximal Abelian links from these configurations are compared to the full SU(2) calculations in Fig. 5-7 for $ma = 0.1$, 0.2 and 0.3 at $\beta = 2.4$. Here we will only be concerned with qualitative features, no mass fits were done. To make the comparison clearer the correlators calculated with Abelian links have been divided by a factor 2.

The shape of the pseudoscalar correlator at all values of $ma$ is remarkably close to the SU(2) calculation, especially at larger times. The splitting between the vector and pseudoscalar correlators, and by inference, the vector-pseudoscalar mass splitting is well reproduced at $ma = 0.1$. However at larger masses the vector-pseudoscalar splitting goes away faster in the Abelian projected calculation than in the full SU(2) calculation. This is qualitatively consistent with what is seen in $< \overline{\chi}\chi > (m)$ where agreement is also better at the smaller masses. It is expected that lighter fermions would be more sensitive to long-distance correlations in the gauge field so an interpretation of our result is that Abelian-projected links in maximal Abelian gauge reasonably describe the long-distance physics of the nonabelian theory but there are some short-distance effects which are not reproduced.

## IV.  SUMMARY

In this work the study of Abelian-projected gauge theory was extended into the light quark sector. Chiral symmetry breaking properties of SU(2) lattice gauge theory were calculated in quenched approximation. Then, after making an Abelian projection in the maximal Abelian gauge and in the field-strength gauge, the chiral order parameter was recalculated using Abelian links in the fermion matrix.

For maximal Abelian gauge projected fields it was found that the expectation value $< \overline{\chi}\chi > (m)$ is a somewhat different function of mass than in the full SU(2) calculations. However the small mass and extrapolated zero mass values are similar. Furthermore, it was found that the chiral order parameter calculated with maximal Abelian links obeys perturbative scaling very well for $2.3 \leq \beta \leq 2.5$.

In contrast, using field strength gauge projected links, behaviour resembling that of a strongly coupled gauge theory was found. The expectation value $< \overline{\chi}\chi > (m)$ was very insensitive to $m$ and $\beta$.

The shape of the pseudoscalar and vector meson correlators calculated with maximal Abelian links was found to agree very well with the $SU(2)$ theory, at least for small fermion mass. At larger masses the splitting between pseudoscalar and vector channels is diminished relative to the full nonabelian calculation.

The results presented here provide an indication that Abelian projected fields can describe the long-distance physics of light quarks in lattice QCD. Of course this description is not unique and many questions about whether the notion of "Abelian dominance", now extended into the quark sector, can be turned into a useful effective theory for QCD remain to be answered.

*Note added:* After submitting this paper we received preprints from Sasaki *et al*[20] and from Miyamura[21] which deal with the role of Abelian monopoles in chiral symmetry breaking of QCD using Schwinger-Dyson equations and lattice techniques respectively. These authors also conclude that chiral symmetry breaking of QCD can be described by an Abelian theory.

## ACKNOWLEDGMENTS

It is a pleasure to thank G. Poulis and H. Trottier for helpful discussions. This work was supported in part by the Natural Sciences and Engineering Research Council of Canada.

## V.   REFERENCES

[1] G. 't Hooft, Proceedings of the EPS International Conference on High Energy Physics, Palermo, 1975, Ed. A. Zichichi (Editorice Compositori, Bologno, 1976); S. Mandelstam, Phys. Rev. 236, 245 (1976).

[2] G. 't Hooft, Nucl. Phys. <u>B190</u>, 455 (1981).

[3] T. Suzuki and I. Yotsuyanagi, Phys. Rev. <u>D42</u>, 4257 (1990); S. Hioki, S. Kitahara, S. Kiura, Y. Matsubara, O. Miyamura, S. Ohno and T. Suzuki, Phys. Lett. <u>B272</u>, 326 (1991).

[4] T.L. Ivaneko, A.V. Pochinsky and M.I. Polikarpov, Phys. Lett. <u>B302</u>, 458 (1993).

[5] L. Del Debbio, A. Di Giacomo and G. Paffuti, Pisa preprint IFUP–TH 16/94 (1994).

[6] H. Shiba and T. Suzuki, Kanagawa preprint 93–09 (1993).

[7] K. Yee, Phys. Rev. <u>D49</u>, 2574 (1994); see also K. Yee, Louisiana State University preprint LSU-0725–94 (1994).

[8] A.S. Kronfeld, G. Schierholz and U.-J. Wiese, Nucl. Phys. <u>B293</u>, 461 (1987).

[9] A.S. Kronfeld, M.L. Laursen, G. Schierholz and U.-J. Wiese, Phys. Lett. <u>B198</u>, 516 (1987).

[10] L. Del Debbio, A. Di Giacomo, M. Maggiore and S. Olejnik, Phys. Lett. <u>B267</u>, 254 (1991).

[11] G.I. Poulis, H.D. Trottier and R.M. Woloshyn, Phys. Rev. D, in press.

[12] V.G. Bornyakov, E.-M. Ilgenfritz, M.L. Laursen, U.K. Mitrjuskin, M. Müller-Preussker, A.J. van der Sijs and A.M. Zadorozhny, Phys. Lett. <u>B261</u>, 116 (1991).

[13] N. Kawamoto and J. Smit, Nucl. Phys. <u>B192</u>, 100 (1981).

[14] R. Scalletar, D. Scalapino and R.L. Sugar, Phys. Rev. <u>B34</u>, 7911 (1986); S. Gottlieb, W. Liu, D. Toussaint, R.L. Renken and R.L. Sugar, Phys. Rev. <u>D35</u>, 3972 (1987); K. Bitar, A.D. Kennedy, R. Horsley, S. Meyer and P. Rossi, Nucl. Phys. <u>B313</u>, 348 (1989).

[15] H.R. Fiebig and R.M. Woloshyn, Phys. Rev. $\underline{D42}$, 3520 (1990).

[16] A. Billoire, R. Lazage, E. Marinari and A. Morel, Nucl. Phys. $\underline{B251}$, 581 (1985).

[17] J. Kogut, M. Stone, H.W. Wyld, S.H. Shenker, J. Shigemitsu and D.K. Sinclair, Nucl. Phys. $\underline{B225}$, 326 (1983).

[18] C. Michael and M. Teper, Nucl. Phys. $\underline{B305}$, 453 (1988).

[19] S.J. Hands and M. Teper, Nucl. Phys. $\underline{B347}$, 819 (1990).

[20] S. Sasaki, H. Suganuma and H. Toki, Riken preprint RIKEN-AF-NP172 (1994).

[21] O. Miyamura, Hiroshima University preprint, to appear in the proceedings of Lattice 94 (Bielefeld, 1994).

Table I. Calculated values of $< \overline{\chi}\chi > (m)$ [Eq. (8)] for SU(2), maximal abelian gauge projected and field-strength gauge projected fields for various values of $\beta$ and fermion mass $ma$. The values are given in lattice units and the numbers in parentheses is the statistical error in the final digit.

| $\beta$ | $ma$ | SU(2) | maximal abelian gauge | field-strength gauge |
|---------|------|-------|-----------------------|----------------------|
| 2.3 | 0.05 | 0.1276(6) | 0.1326(9) | 0.543(1) |
| | 0.10 | 0.1828(4) | 0.1704(7) | 0.5428(5) |
| | 0.15 | 0.2264(3) | 0.2038(7) | 0.5418(5) |
| | 0.20 | 0.2621(4) | 0.2332(6) | 0.5402(4) |
| | 0.25 | 0.2916(3) | 0.2592(5) | 0.5372(3) |
| | 0.30 | 0.3161(2) | 0.2816(5) | 0.5348(4) |
| 2.35 | 0.05 | 0.1070(4) | 0.1085(8) | 0.5394(6) |
| | 0.10 | 0.1635(3) | 0.1482(7) | 0.5390(7) |
| | 0.15 | 0.2091(3) | 0.1823(6) | 0.5376(4) |
| | 0.20 | 0.2469(3) | 0.2136(6) | 0.5372(5) |
| | 0.25 | 0.2778(2) | 0.2406(5) | 0.5350(3) |
| | 0.30 | 0.3039(2) | 0.2650(5) | 0.5324(2) |
| 2.40 | 0.05 | 0.0909(5) | 0.0867(9) | 0.5362(8) |
| | 0.10 | 0.1482(4) | 0.1272(8) | 0.5370(6) |
| | 0.15 | 0.1953(3) | 0.1694(7) | 0.5362(4) |
| | 0.20 | 0.2341(3) | 0.1948(7) | 0.5352(5) |
| | 0.25 | 0.2665(3) | 0.2236(5) | 0.5332(4) |
| | 0.30 | 0.2937(3) | 0.2490(5) | 0.5340(4) |
| 2.45 | 0.05 | 0.0807(4) | 0.0738(9) | 0.5350(8) |
| | 0.10 | 0.1375(3) | 0.1140(8) | 0.5354(6) |
| | 0.15 | 0.1846(3) | 0.1498(7) | 0.5336(6) |
| | 0.20 | 0.2244(3) | 0.1826(6) | 0.5334(4) |
| | 0.25 | 0.2576(3) | 0.2120(6) | 0.5318(4) |
| | 0.30 | 0.2857(3) | 0.2380(5) | 0.5292(3) |
| 2.50 | 0.05 | 0.0718(3) | 0.0623(5) | 0.5326(7) |
| | 0.10 | 0.1281(2) | 0.1018(5) | 0.5330(6) |
| | 0.15 | 0.1755(2) | 0.1379(5) | 0.5332(4) |
| | 0.20 | 0.2154(2) | 0.1710(4) | 0.5318(4) |
| | 0.25 | 0.2494(2) | 0.2001(4) | 0.5298(4) |
| | 0.30 | 0.2781(2) | 0.2284(3) | 0.5278(4) |

Table II. Values of the chiral condensate $< \overline{\chi} \chi >_0$ calculated in SU(2) and maximal Abelian projected fields. Numbers in parenthesis are estimated errors in the final digit.

| $\beta$ | SU(2) | maximal Abelian gauge |
|---|---|---|
| | 6-point fit | |
| 2.3 | 0.071(1) | 0.092(1) |
| 2.35 | 0.048(1) | 0.067(1) |
| 2.4 | 0.031(1) | 0.044(1) |
| 2.45 | 0.020(1) | 0.031(2) |
| 2.5 | 0.0122(5) | 0.020(1) |
| | 5-point fit | |
| 2.3 | 0.079(2) | 0.093(3) |
| 2.35 | 0.056(1) | 0.069(3) |
| 2.4 | 0.037(1) | 0.046(3) |
| 2.45 | 0.026(1) | 0.032(3) |
| 2.5 | 0.0165(7) | 0.021(2) |

## Figure Captions

1. The expectation value of staggered fermion fields $< \overline{\chi}\chi >$ as a function of fermion mass in lattice units at $\beta = 2.3$ for the SU(2) theory ($\Delta$), maximal Abelian gauge projected fields ($\circ$) and field-strength gauge projected fields ($\square$). Also shown are the fits of the extrapolation Eq. (10) to the SU(2) and maximal Abelian results.

2. The expectation value of staggered fermion fields $< \overline{\chi}\chi >$ as a function of fermion mass in lattice units at $\beta = 2.5$ for the SU(2) theory ($\Delta$), maximal Abelian gauge projected fields ($\circ$) and field-strength gauge projected fields ($\square$). Also shown are the fits of the extrapolation function Eq. (10) to the SU(2) and maximal Abelian results.

3. Check for perturbative scaling in the chiral condensate calculated with SU(2) links (a) and with maximal Abelian gauge projection (b). The squares are the results of the 6-point fit to Eq. (10) and the circles, the 5-point fit.

4. Check for scaling of the chiral condensate, calculated with SU(2) links (a) and with maximal Abelian gauge projection (b), relative to the string tension, $r_{st}(\beta) = < \overline{\chi}\chi >_0^{\frac{1}{3}} / \sqrt{K}$. The squares are the results of the 6-point fit to Eq. (10) and the circles, the 5-point fit.

5. The pseudoscalar (squares) and vector (circles) meson correlators as a function of Euclidean time and calculated at $\beta = 2.4$ and $ma = 0.1$. The open symbols are for the SU(2) theory, the filled symbols are for maximal Abelian gauge projection with the correlators divided by 2.

6. Same as Fig. 5 at $ma = 0.2$.

7. Same as Fig. 5 at $ma = 0.3$.

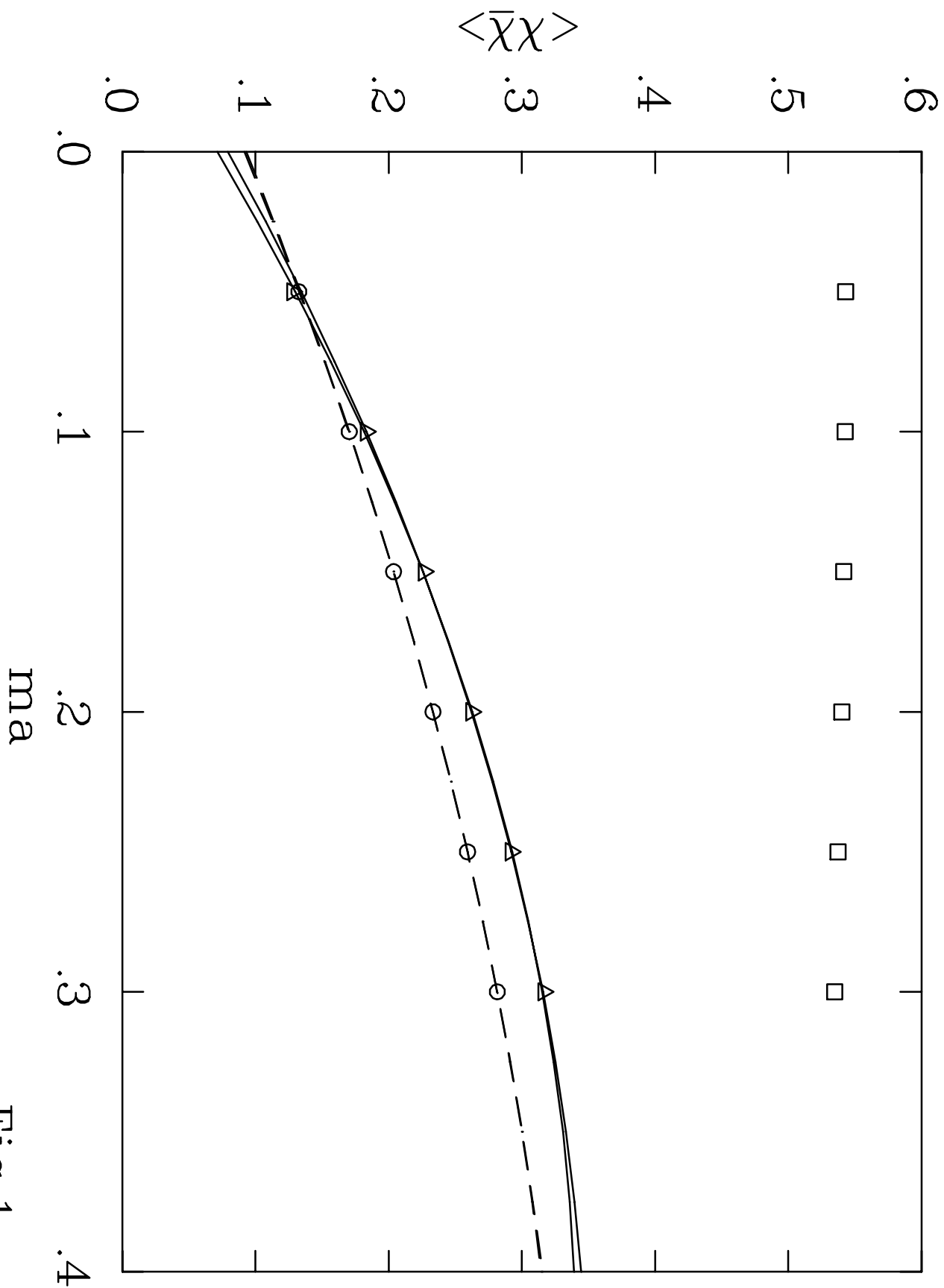

Fig.1

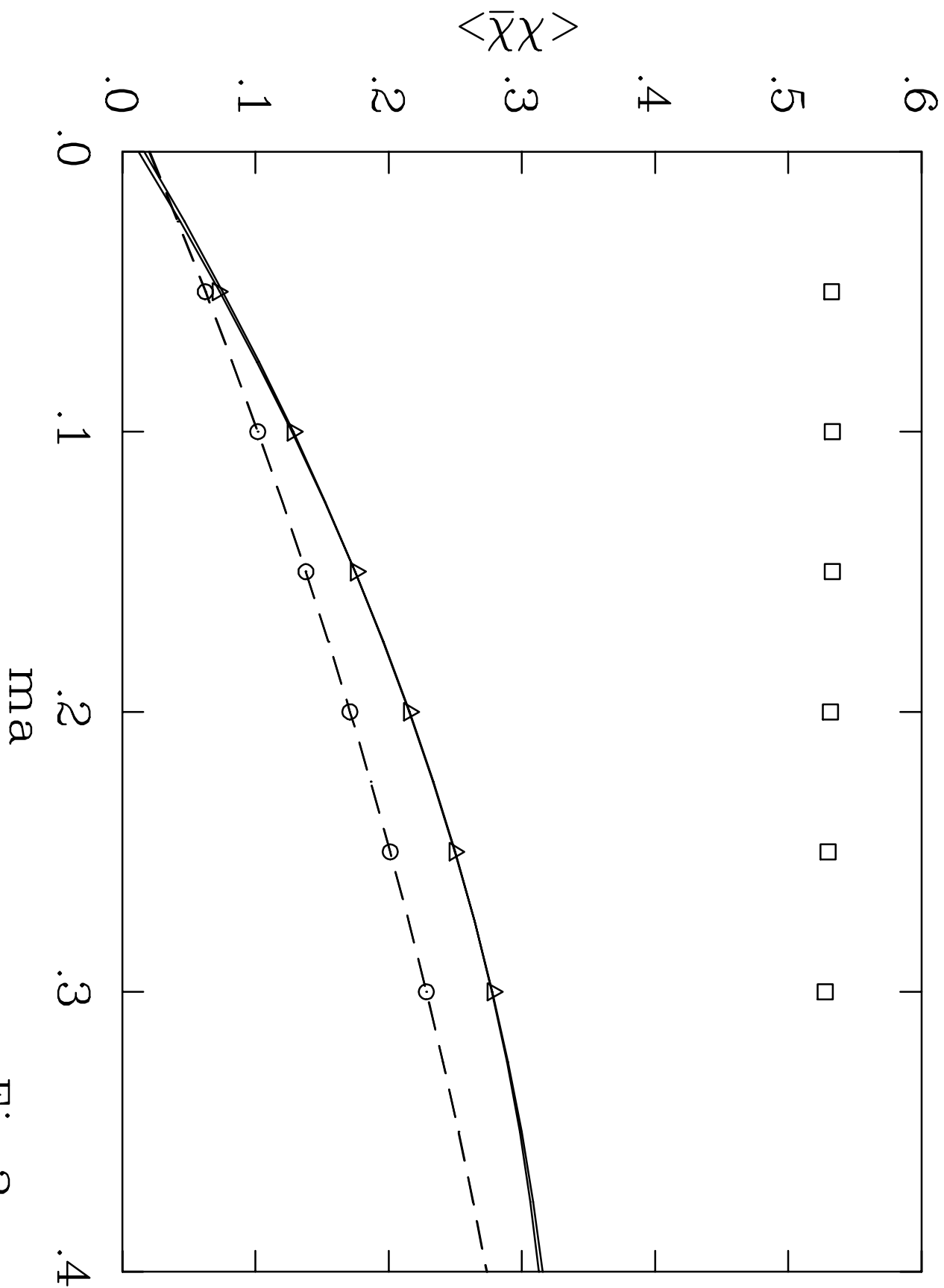

Fig.2

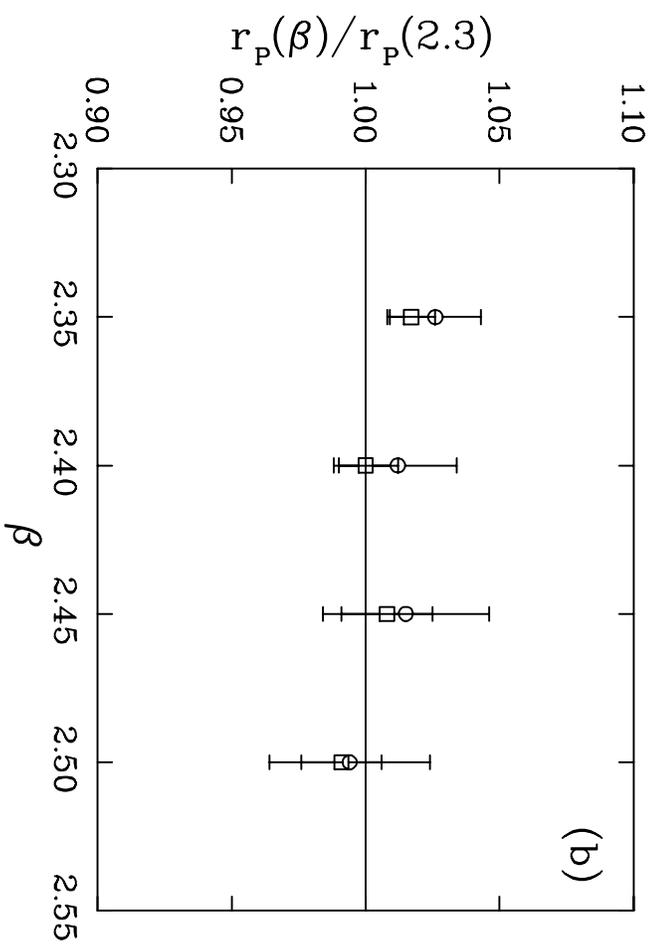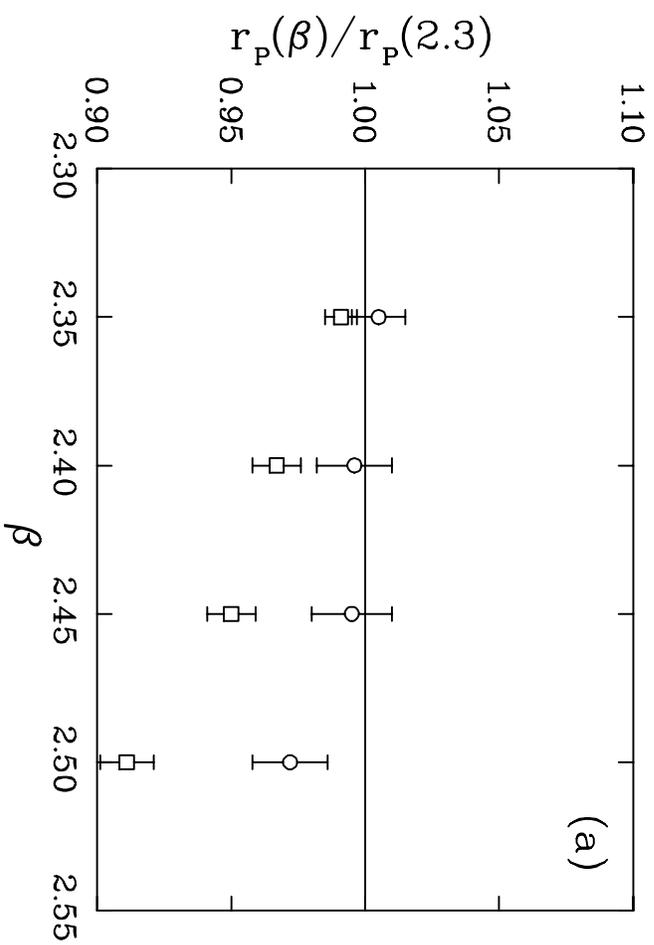

Fig. 3

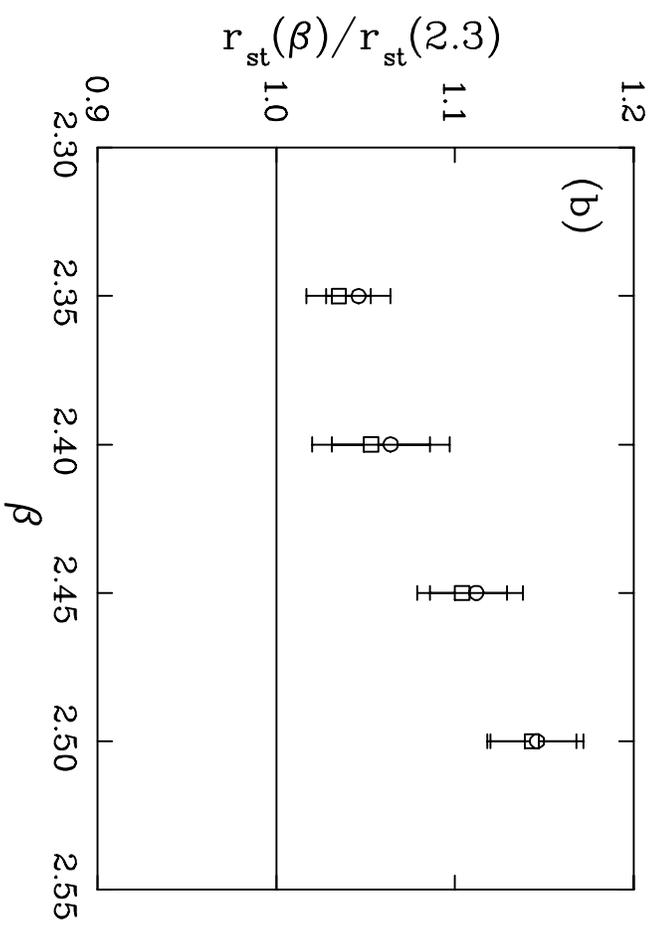
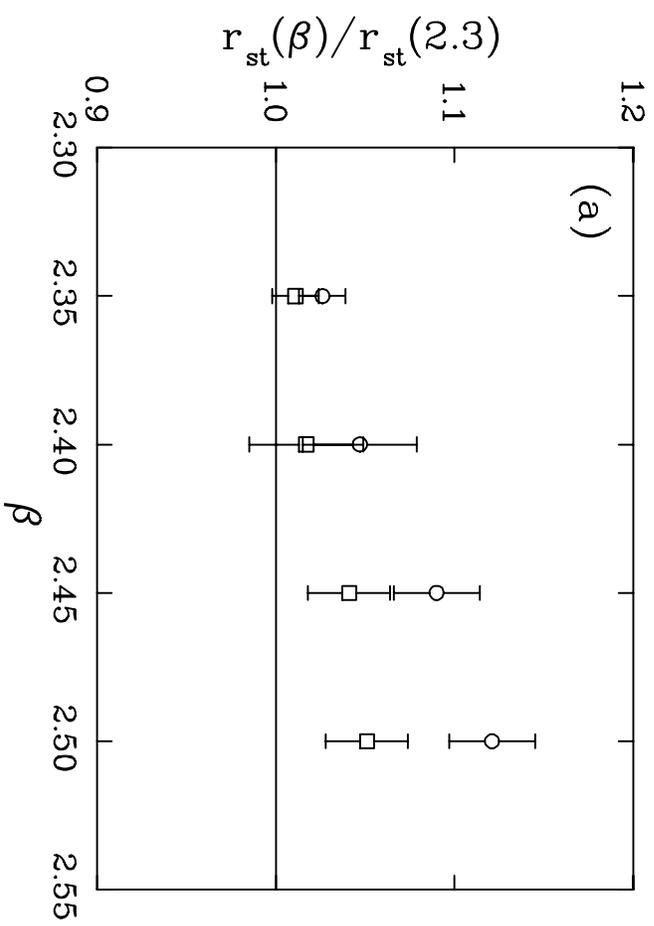

Fig. 4

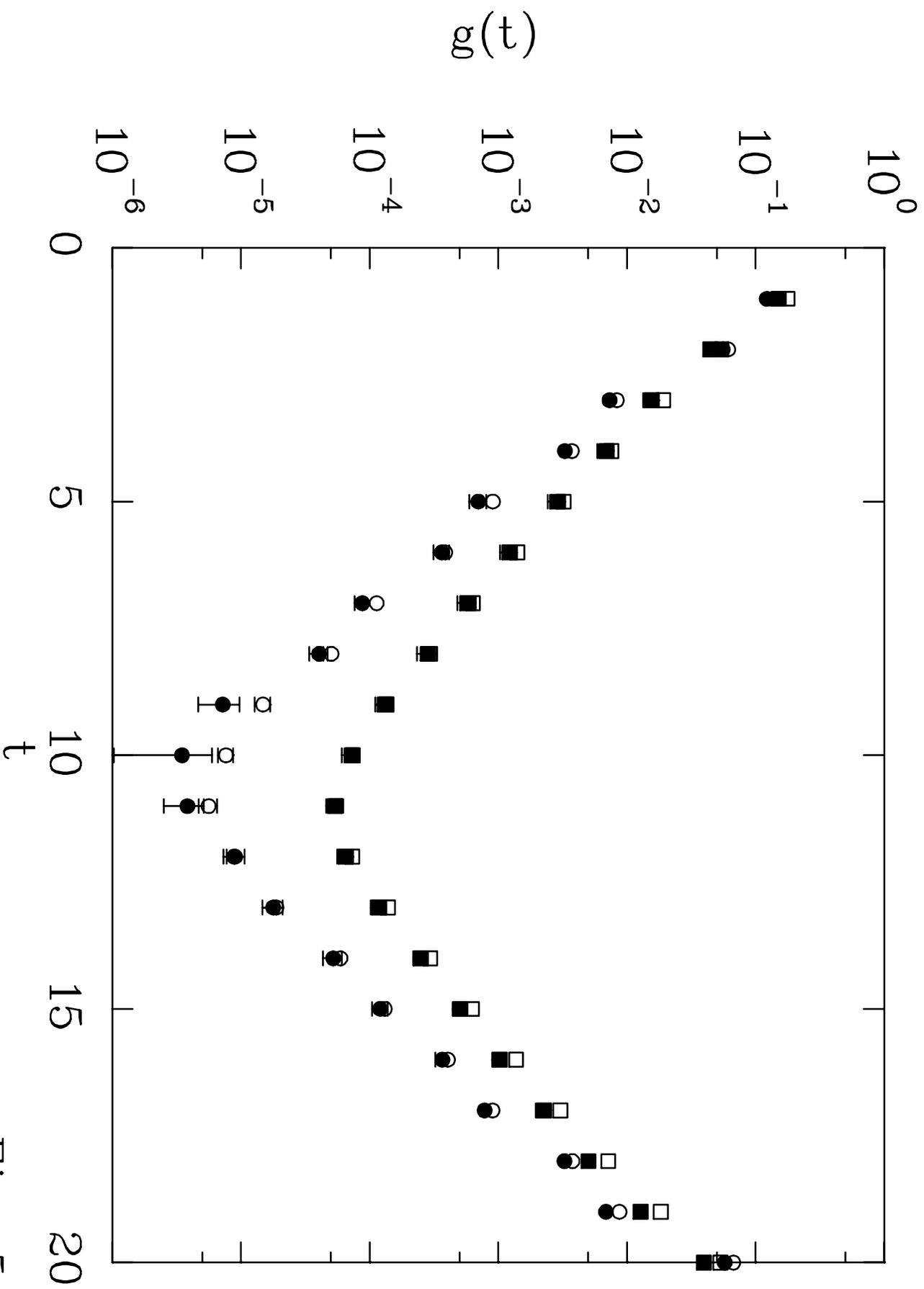

Fig. 5

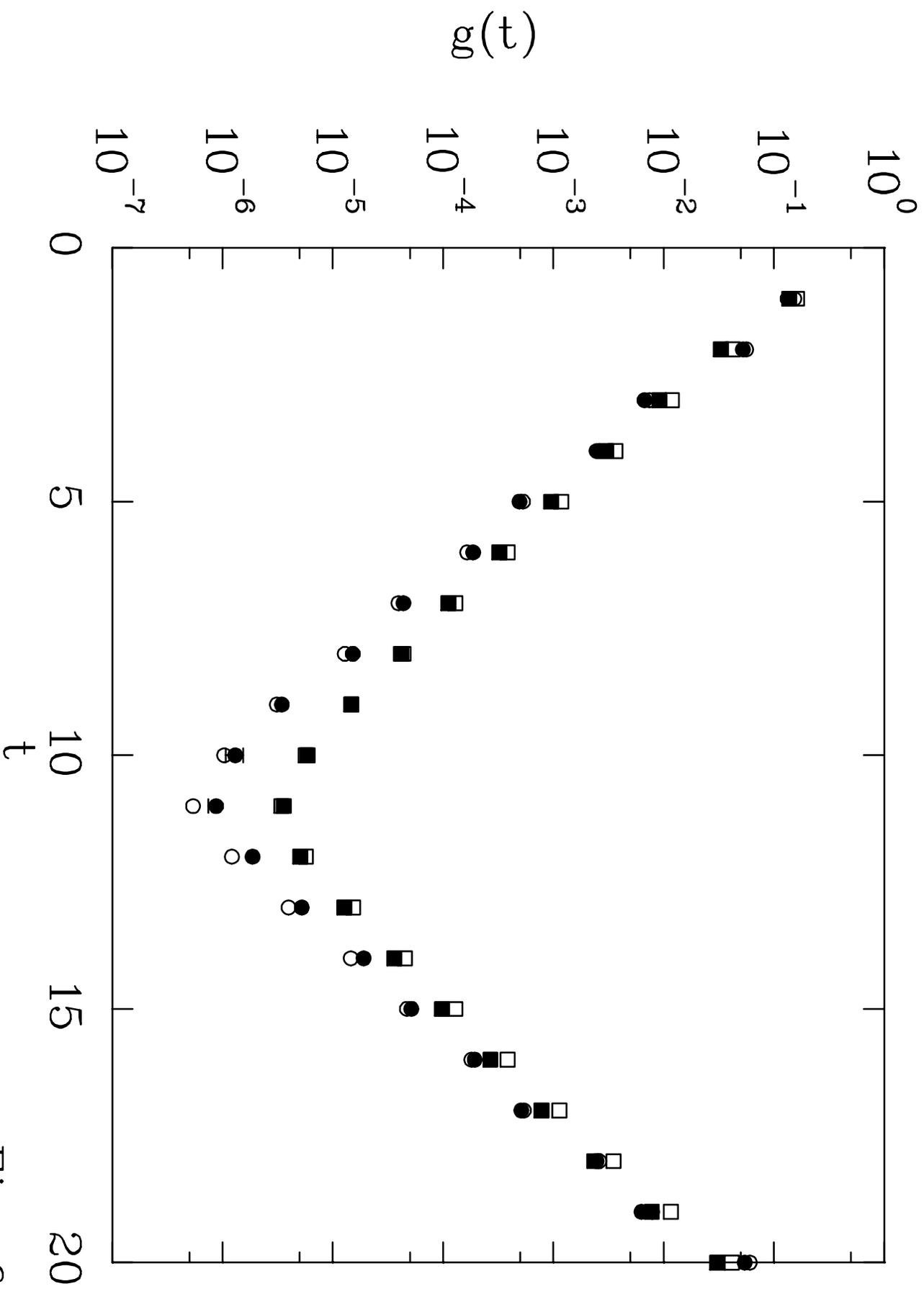

Fig. 6

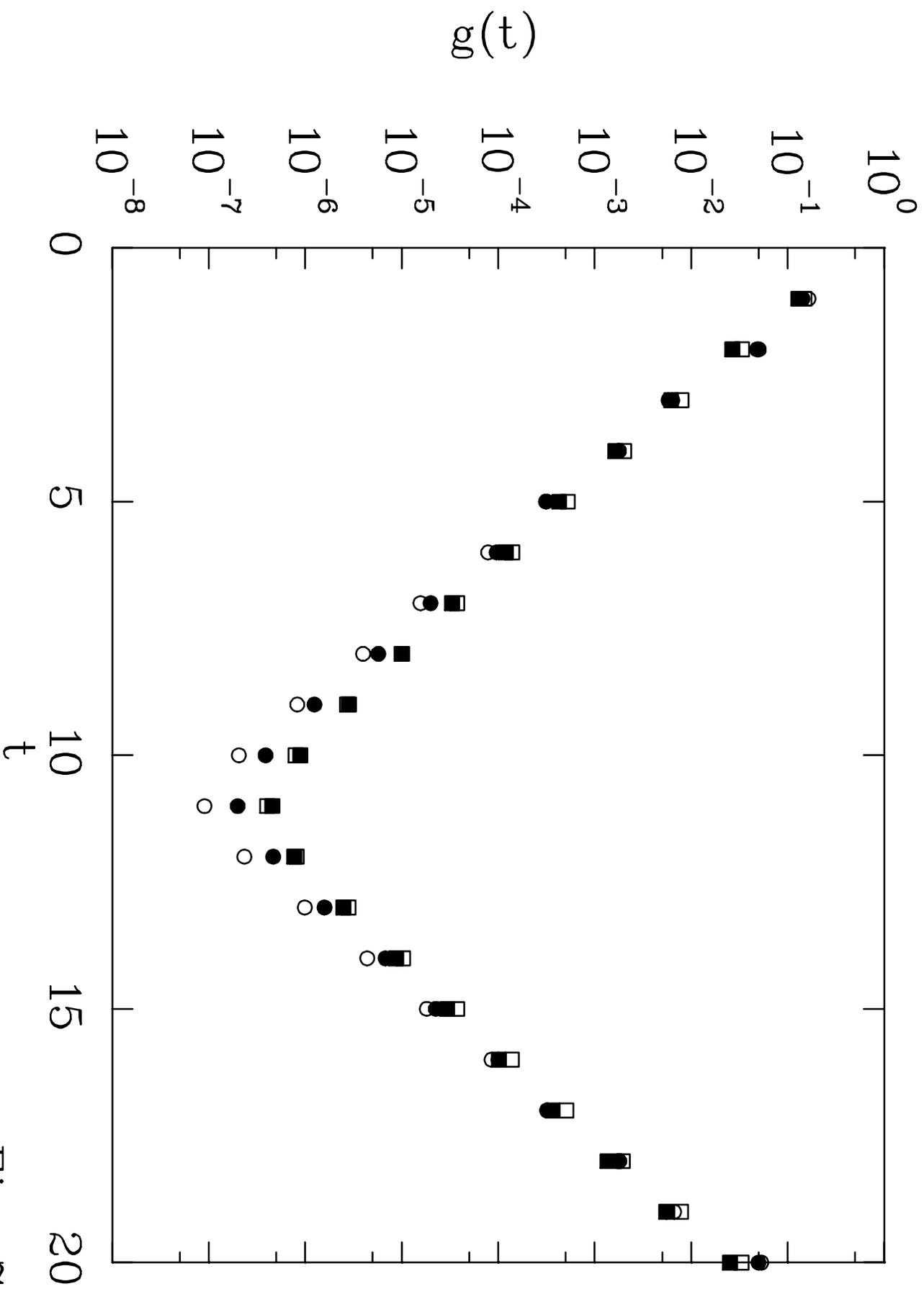

Fig. 7